\documentclass[sigconf, nonacm]{acmart}
\AtBeginDocument{%
  }

\setcopyright{acmlicensed}
\copyrightyear{2018}
\acmYear{2018}
\acmDOI{XXXXXXX.XXXXXXX}
\acmConference[Conference acronym 'XX]{Make sure to enter the correct
  conference title from your rights confirmation email}{June 03--05,
  2018}{Woodstock, NY}
\acmISBN{978-1-4503-XXXX-X/2018/06}

\usepackage{longtable}
\usepackage{tabularx}
\usepackage{stfloats}

\begin{document}

\title{A Scale For Value Alignment In Human-AI Interaction}

\author{Lena Hegemann}
\affiliation{%
  \institution{Hasso Plattner Institute}
  \city{Potsdam}
  \country{Germany}
}
\email{lena.hegemann@hpi.de}

\author{Steeven Villa}
\affiliation{%
  \institution{LMU Munich}
  \city{Munich}
  \country{Germany}}
\email{steeven.villa@ifi.lmu.de}

\author{Hyemin Bang}
\affiliation{
    \institution{MIT}
    \city{Cambridge}
    \state{Massachusetts}
    \country{USA}
}
\email{hbang@mit.edu}

\author{Mitchell L. Gordon}
\affiliation{
    \institution{MIT}
    \city{Cambridge}
    \state{Massachusetts}
    \country{USA}
}
\email{mlgordon@mit.edu}

\author{Antti Oulasvirta}
 \affiliation{%
  \institution{Aalto University\\ ELLIS Institute Finland}
  \city{Espoo}
  \country{Finland}
 }
\email{antti.oulasvirta@aalto.fi}

\author{Robin Welsch}
 \affiliation{%
  \institution{Aalto University}
  \city{Espoo}
  \country{Finland}
 }
\email{robin.welsch@aalto.fi}

\author{Patrick Ebel}
\affiliation{%
  \institution{Hasso Plattner Institute}
  \city{Potsdam}
  \country{Germany}
}
\email{patrick.ebel@hpi.de}

\renewcommand{\shortauthors}{Hegemann et al.}

\newcommand{\factora}{Value Understanding}
\newcommand{\fa}{VU}
\newcommand{\factorb}{Value Manifestation}
\newcommand{\fb}{VM}
\newcommand{\scaleshort}{PVAS}

\begin{abstract}
\emph{Value alignment} is a central objective in AI and HCI research, yet no validated instrument measures how users perceive it. This gap hampers the comparison and accumulation of findings and limits the effectiveness of applications, where understanding users' viewpoints is critical. We construct and evaluate a 13-item psychometric scale that measures perceived value alignment across two components: value understanding and value manifestation. It is based on a large item pool drawn from prior empirical studies, filtered by experts, and finally assessed by users (N=607) across diverse AI scenarios. Confirmatory factor analysis on an independent sample (N=259) confirmed the two-factor structure and high internal consistency for both subscales. Using optimization, we also derived a 6-item short form for quick administration. The scale is a reliable measure, providing HCI researchers with a common evaluation metric across contexts.

\end{abstract}


\begin{teaserfigure}
 \includegraphics[width=\textwidth]{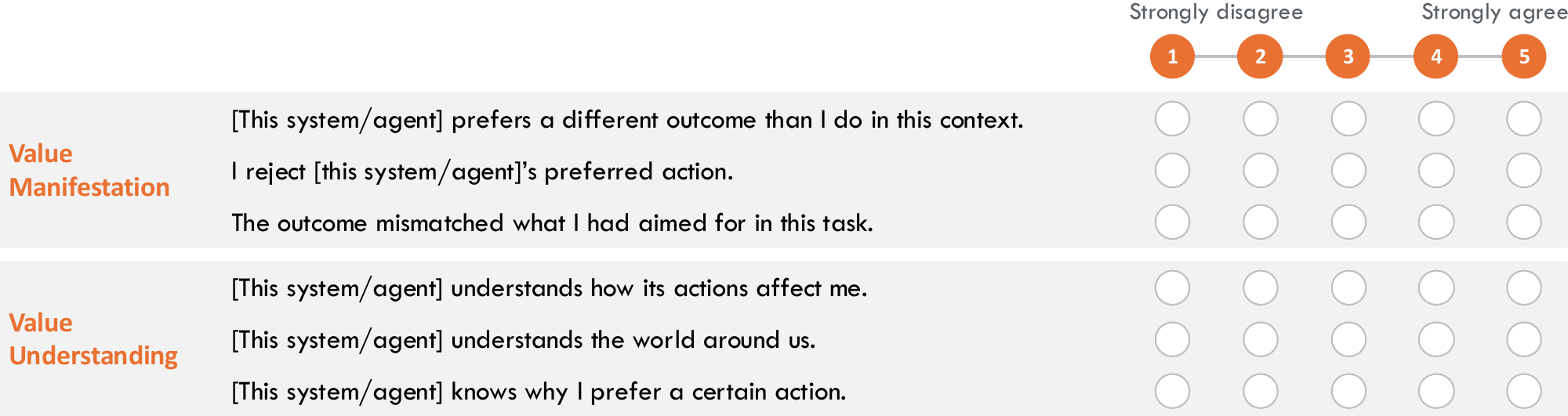}
  \caption{The six-item short form of the Perceived Value Alignment Scale.}
  \Description{Enjoying the baseball game from the third-base
  seats. Ichiro Suzuki preparing to bat.}
  \label{fig:teaser}
\end{teaserfigure}

\maketitle

\section{Introduction}

Suppose a digital assistant handles your correspondence while you are away. A student asks for a thesis extension after a family emergency. Because you value consistency and have declined such requests before, the assistant sends a polite refusal in your name, citing your rules. Yet you would have granted this extension, at this severity, care for the student outweighs the rule. The wrong email has been sent by an assistant that had your information and your precedents, but not the values behind them. To design for alignment, or to reduce misalignment of this kind, it must first be measurable.

The above example illustrates \textit{value alignment}: the problem of ensuring that an AI system's behavior is congruent with what its users consider important~\cite{smart_theory_2026, mckinlay_understanding_2026}.
As intelligent interactive systems grow more autonomous, often acting without constant human approval, the problem is becoming an increasingly important one~\cite{chen_code_2026, he_planthenexecute_2025}.
These systems build persistent models of their users~\cite{shaikh_creating_2025}, make proactive suggestions~\cite{susak_provoice_2026, chen_need_2025}, and act on users' behalf~\cite{zhang_characterizing_2026}.

However, while value alignment has originated from AI research, HCI research has a  central role in tackling the problem. Alignment is not only a property of a model, but it is also an essential emergent consequence of interaction~\cite{shen_bidirectional_2025, fan_userdriven_2025}.
How a system elicits what matters to its user~\cite{yun_ai_2026}, how it presents the reasons for its actions, and when it asks rather than acts~\cite{sim_words_2026} all shape whether users perceive its behavior as aligned.
In current systems, this is ensured during the interaction itself: users prompt a system for one task at a time, inspect each result, and can reject or revise it.

The increasing agency and generality of systems changes what should be evaluated in human--AI interaction.
AI systems will act in situations that no study can realistically cover, so task-level measures say little about behavior beyond what was tested.
However, the constructs HCI relies on are of this kind.
Trust~\cite{lee_trust_2004, vereschak_how_2021}, human--AI teaming~\cite{oneill_human_2022, seeber_machines_2020}, and shared mental models~\cite{mathieu_influence_2000, bansal_accuracy_2019} capture whether a system can be relied on for a task, whether the team performs on it, and whether both hold compatible representations of it.
What they do not capture is whether the human and the system agree on what is worth pursuing.
An agent can be trusted, team effectively, and share a task model, and still pursue outcomes its user would not endorse, as in the opening example.
Whether users perceive an agent as acting in line with their values generalizes beyond the tested situations, and therefore complements task-level evaluation for systems that act across tasks and contexts.

So far, researchers have had no reliable way to measure whether users perceive an AI system as aligned with their values.
HCI has already begun trying to measure value alignment, but without a common instrument.
Our literature analysis (see \autoref{sec:item_generation} showed that recent user studies evaluate aspects of ``\textit{alignment}'' in various senses, yet rely on self-developed, non-validated scales~\cite{hegemann_tomigo_2026, kim_anthropomorphic_2023, zheng_chat_2026}.
There is also no instrument to adopt from AI research, either.
Although aligning models with human values has driven research and industry for years~\cite{sorensen_roadmap_2024}, the community's understanding of values is often ill-defined~\cite{smart_theory_2026}, and neither a common understanding of the construct nor an established way to measure it exists.
Neither field, therefore, offers an instrument whose scores can be interpreted as perceived value alignment.
The measures in use leave open whether they are sufficient, capture alignment or an adjacent construct such as trust or teaming, and whether their scores are reliable.
Since every study measures differently, it is equally unknown whether results can be compared.

A validated instrument is therefore needed now, while evaluation practices for interactive AI are still forming.
We argue that such an instrument must generalize across tasks and environments and apply to any human--AI interaction in which a user can observe an agent's behavior.
Because users can judge an agent only by the behavior they observe, we develop a scale to assess \textit{perceived} value alignment.
A system's internal objectives are not accessible to its users, and it is their perception that shapes the interaction between human and AI.

To develop the Perceived Value Alignment Scale, we followed established best practices for scale development~\cite{boateng_best_2018}, which separate item development, scale development, and scale evaluation.
Following the deductive method, we generated the initial item pool from items and constructs used in prior HCI work to assess related aspects of human--AI interaction, such as teaming~\cite{attig_measuring_2026} and shared understanding, thereby grounding the scale in the accumulated knowledge of the domain rather than in ad hoc formulations.
In an expert study, we selected the 106 items with the highest content validity.
607 users then rated 6 scenarios that varied in alignment level, use case, and interaction paradigm.
Exploratory factor analysis revealed a 2-factor structure comprising value understanding and value manifestation.
We tested the scale on an independent sample (N259) using confirmatory factor analysis.
The resulting subscales showed high internal consistency, convergent, discriminant and known group validity.

The scale allows researchers to evaluate perceived value alignment as a dependent variable in user studies and to compare systems and design variants.
In summary, we contribute (1) a theoretically grounded definition of perceived value alignment of interactive AI, (2) a validated psychometric scale, and (3) evidence for its factorial validity and internal consistency.

\section{Related Work}

\begin{figure*}
    \centering
    \includegraphics[width=\textwidth]{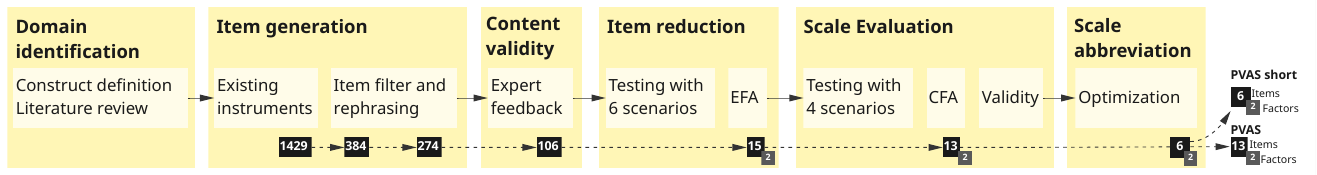}
    \caption{Overview of the scale development process following \citet{boateng_best_2018}. The numbers indicate the number of items and factors at each stage. The resulting scale exists in a full and short version.}
    \label{fig:process}
    \Description{Diagram illustrating the scale development process adapted from Boateng et al., divided into three main phases. Stage one, Item Generation, shows domain identification leading to 1,429 items from existing literature, which were filtered and rephrased down to 274 items. Stage two, Content Validity, shows item reduction using expert feedback to select 106 items. Stage three, Scale Evaluation, shows user testing across scenarios analyzed with Exploratory Factor Analysis to yield a 15-item model, followed by Confirmatory Factor Analysis and validity testing to produce the final 13-item PVAS, which was then shortened via optimization to form the 6-item PVAS short form.}
\end{figure*}
Work relevant to perceived value alignment spans three strands: validated HCI scales for adjacent constructs, evaluation practices for AI value alignment, and HCI studies that assess alignment in specific interactions. In the following, we review each. 

\subsection{Psychometric Scales}

HCI offers established scales for assessing perceptions and dispositions relevant to human--AI interaction.
The Trust in Automation questionnaire assesses trust and related perceptions, including reliability, competence, and predictability~\cite{körber_theoretical_2019}.
These perceptions may overlap with perceived value alignment, but confidence in a system's capabilities does not by itself establish whether its behavior reflects what matters to its user.
Usability instruments, including the Usefulness, Satisfaction, and Ease of Use (USE) questionnaire~\cite{gao_psychometric_2018} and the System Usability Scale (SUS)~\cite{lewis_system_2018}, assess qualities of using a system; a system may be easy to use while pursuing outcomes that conflict with the user's priorities.

Teaming and cooperativity scales focus on how humans and AI work together.
Attig et al.'s Perceived Cooperativity Scale and Teaming Perception Scale assess an agent's perceived cooperative capability and practice, and the sense of working together through mutual contribution and support, respectively~\cite{attig_measuring_2026}.
The Human-Machine-Interaction-Interdependence Questionnaire similarly assesses perceived interdependence in driver--vehicle cooperation~\cite{woide_measuring_2021}.
While it is plausible that alignment supports a system's ability to cooperate with humans and work with them in teams, perceived value alignment should also apply when users observe an agent's behavior without directly cooperating with it.

Other scales address users' affective responses or general dispositions toward technology.
The Perceived Creepiness of Technology Scale includes implied malice, undesirability, and unpredictability~\cite{woźniak_creepy_2021}.
Creepiness is a plausible consequence of misalignment, but not all misalignment implies malicious intentions by developers or makes a system less predictable.
An agent may consistently pursue outcomes that conflict with a user's values without being perceived as creepy.
The Affinity for Technology Interaction (ATI) scale and its short form assess a person's general tendency to engage with technology~\cite{franke_personal_2019,wessel_atis_2019}.
They therefore characterize the user rather than their perception of alignment with a particular system after interaction.

Perceived shared understanding is more closely related to our construct.
Liang and Banks' validated scale assesses users' sense of sharing meaning with an AI system~\cite{liang_perceived_2025}.
This overlaps with the understanding component of perceived value alignment, but a shared interpretation of a situation does not necessarily imply that the system's actions reflect the user's priorities.
More directly, Liu et al. develop and validate the Human-AI Companion Alignment (HACA) scale, which captures cognitive, affective, and behavioral alignment as aspects of companion personalization~\cite{liu_thank_2026}.
Its dimensions address agreement with users' opinions, emotional responsiveness, and similarity in communication style.
The scale's applicability beyond AI-companion relationships remains to be established.

Our work builds on these instruments to provide \scaleshort{}, a scale targeting perceived congruence in values, goals, and preferences through judgments of an agent's observed behavior.

\subsection{AI Value Alignment: Concepts and Evaluation}
\label{sec:alignment-signals}

For HCI, value alignment raises the question of whether an AI system's behavior reflects what matters to its users~\cite{mckinlay_understanding_2026}.
AI alignment is assessed through several kinds of evidence, from preferences over model outputs to judgments formed during interaction.
In language-model training, reinforcement learning from human feedback uses human comparisons to fit a reward model, whereas Direct Preference Optimization learns directly from preferred--dispreferred response pairs~\cite{ouyang_training_2022,rafailov_direct_2023}.
These comparisons operate at the preference level of the values--goals--preferences distinction introduced above.
At evaluation time, Chatbot Arena similarly aggregates anonymous pairwise preference votes to compare models~\cite{chiang_chatbot_2024}; LLM-as-a-judge methods automate output assessment, although judgments can be sensitive to response order and verbosity~\cite{zheng_judging_2023}.
Criterion-based evaluations instead assess responses against predefined standards, such as helpfulness, honesty, and harmlessness or the moral judgments encoded in ETHICS~\cite{askell_general_2021,hendrycks_aligning_2021}.
These approaches assess preferred outputs or performance against specified criteria; they do not directly measure a particular user's perception that an agent's behavior reflects their values, goals, and preferences.

Other approaches compare model responses with human opinion or value profiles.
OpinionQA measures correspondence between a model's response distribution and survey-response distributions from demographic groups~\cite{santurkar_whose_2023}.
Norhashim and Hahn combine GPT-3.5's agreement with dataset-designated normative actions in Moral Stories with consistency across repeated judgments, then map the dataset's norms to a human-value taxonomy~\cite{norhashim_measuring_2024}.
Such profiles characterize agreement with a selected reference, but stated positions need not predict contextual choices: ValueActionLens identifies discrepancies between models' stated values and their selections among value-informed actions~\cite{shen_mind_2025}.
Neither population-level correspondence nor consistency with a normative dataset establishes how an interacting user experiences the system's alignment with their own values.

The interpretation of these evaluations depends on what is meant by alignment.
In their review of 94 AI value-alignment papers, Smart et al. found that 79\% neither defined nor described values and 82\% treated preferences as a stand-in for values~\cite{smart_theory_2026}.
This finding describes a pattern in their sample, rather than every approach reviewed here: ETHICS specifies moral criteria, while Norhashim and Hahn explicitly map judgments to a human-value taxonomy~\cite{hendrycks_aligning_2021,norhashim_measuring_2024}.
Preference judgments can provide evidence about values, but the two are not interchangeable: context-specific choices do not exhaust the meaning or normative significance of what people value~\cite{zhi_xuan_beyond_2024}.
Following the measurement perspective of Wallach et al., interpreting a score as alignment requires an explicit construct and evidence supporting that interpretation~\cite{wallach_evaluating_2025}.
This requirement applies equally to behavioral indicators and self-report instruments used to assess alignment in interaction.

\subsection{Perceived Value Alignment in Human--AI Interaction}
\label{sec:perceived-value-alignment}

HCI research provides evidence at the level of interaction.
Studies of AI-assisted programming combine suggestion-acceptance logs with retrospective labels of activities such as evaluating, verifying, and editing suggestions~\cite{mozannar_reading_2024}.
Decision-support studies measure whether a user's final decision agrees with the AI's suggestion and examine team performance separately for correct and incorrect AI recommendations~\cite{bansal_does_2021}.
These observations capture what users do, but cannot by themselves establish why users accept an output or whether it reflects what matters to them.

Creative systems also elicit subjective judgments: ToMigo~\cite{hegemann_tomigo_2026}, PromptCharm~\cite{wang_promptcharm_2024}, DuetUI~\cite{xu_duetui_2026} use Likert ratings of how well generated designs and design concepts align with creative intent.
These ratings address perceived fit within a particular creative task; they do not establish a validated scale of perceived value alignment. Similarly, \citet{yang_who_2026} use unvalidated questionnaires specifically designed to assess relationships between AI roles and trust. 

In online decision-making, Zheng et al. examine alignment between user requirements and AI support through ContrXAI, a mixed-initiative system evaluated as a technology probe~\cite{zheng_chat_2026}.
More directly, Yun et al.'s Value-Alignment Perception Toolkit (VAPT) offers a reusable, modality-agnostic methodology for examining how users perceive AI's extraction, embodiment, and explanation of their values~\cite{yun_ai_2026}.
Their chatbot study combines interactive probes, interviews, and ratings with the validated PVQ-RR questionnaire as a baseline for participants' values.
VAPT therefore already addresses perceived value alignment; its contribution is a probe-based evaluation methodology, not a psychometrically validated self-report scale of that perception.

Taken together, these studies demonstrate a recurring need to assess whether AI systems understand and act on what matters to users.
Existing instruments address neighboring constructs or particular relationships, while task-specific ratings and situated methodologies provide complementary forms of evidence.
Our work develops a common self-report instrument explicitly grounded in perceived congruence in values, goals, and preferences.
\scaleshort{} distinguishes \factora{}, concerning whether a system appears to understand what matters to its user, from \factorb{}, concerning whether its actions and outcomes reflect those values.
We formulate items for use across agent roles and interaction modalities, and evaluate the scale using observed scenarios that vary in embodiment, autonomy, and the values at stake.
Through these scenario-based studies, we examine its factor structure, reliability, and relationships with established measures of trust, creepiness, and affinity for technology interaction.
This provides a basis for assessing perceived understanding and its manifestation in behavior separately within human--AI interaction studies.

\section{Method}

We followed the scale development process and best-practice recommendations proposed by \citet{boateng_best_2018}. This framework structures scale development into a sequence of stages covering construct definition, item generation and refinement, assessment of the scale's latent structure, and evaluation of its psychometric properties. Accordingly, \autoref{sec:definition} through \autoref{sec:evaluation} describe each stage of our scale development process and the corresponding methodological decisions and analyses. After establishing the psychometric properties of the resulting scale, we further examined whether its length could be reduced while preserving its predictive performance. In \autoref{sec:short-form}, we therefore apply the scale-shortening approach proposed by \citet{yarkoni_abbreviation_2010} to derive a more parsimonious version of the instrument.

\section{Construct Definition}
\label{sec:definition}

Scale development begins with specifying the construct to be measured~\cite{boateng_best_2018}.
The definition delimits what the instrument is intended to cover, guides the generation of items, and serves as the reference against which their content validity is judged.
Without it, items can only be assessed against an implicit notion of the construct, and the resulting scores cannot be interpreted~\cite{wallach_evaluating_2025}.

Following \citet{mckinlay_understanding_2026}, we define perceived value alignment as the extent to which a human perceives congruence of values, goals, and preferences with an artificial agent after interacting with it and observing its behavior.
The definition rests on a three-level hierarchy:

\emph{Values.} Adopted from \citet{schwartz_measuring_2022}, values are trans-situational, desirable end states or modes of conduct that vary in importance and serve as guiding principles for a human or artificial agent. They are abstract and motivational, organized as a system of relative priorities rather than a set of independent endorsements, and become observable only through their expression in specific contexts.

\emph{Goals.} Goals are situated manifestations of values: concrete, context-bound end states an actor pursues within a given task, constructed in service of one or more underlying values, where the originating values are not necessarily recoverable from the goal alone (\cite{mckinlay_understanding_2026}; consistent with goal-systems theory~\cite{kruglanski2018theory}).

\emph{Preferences.} Preferences are comparative priorities among options, outcomes, or courses of action given an actor's goals and values (\cite{mckinlay_understanding_2026}; in the decision-theoretic sense of preference as a comparative ranking~\cite{peterson2017introduction}).

Following \citet{mckinlay_understanding_2026}, value alignment is the process of understanding and calibrating values to achieve congruence between two agents. Perceived value alignment, then, is the extent to which a human perceives congruence based on their observation of the agent.

Since values become observable only through their expression in specific contexts, the construct is assessed after interaction with an intelligent system.

\section{Item Generation}
\label{sec:item_generation}

We generated the initial item pool in three steps. First, we identified
questionnaires measuring constructs related to perceived value
alignment through a literature analysis and extracted 1429 items reported in these instruments. Second, we generalized and
unified the phrasing of the items so that they no longer referred to a
particular system, task, outcome, or AI role. Third, we removed
redundant, irrelevant, and out-of-scope items. The wording of the remaining 274 items was revised for linguistic naturalness.

\paragraph{Literature Analysis.} We approached perceived value alignment through a set of adjacent constructs that describe how people interact with other people or artificial agents, where value alignment is essential. Based on our initial literature review and construct definition, three of the authors collaboratively identified the following keywords: \emph{collaboration}, \emph{shared understanding},
\emph{cooperation}/\emph{cooperativity}, \emph{teaming}, \emph{value
alignment}, \emph{value-action gap}, \emph{cohesion}, \emph{partnership},
\emph{integration}, and \emph{symbiosis}. For each keyword, we queried
\textit{Scopus}\footnote{https://www.scopus.com/, accessed on Sep 3, 2026} in full-text mode following the pattern
\texttt{"Human-AI" AND "<key word>" AND "questionnaire"}. The searches were conducted in July 2026. For search queries with more than 100 results, only the 50 newest and 50 most-cited results were viewed, to balance the representation of the different constructs. From the resulting publications, 79 used a unique questionnaire targeting a relevant construct. The retrieved papers were screened for questionnaires measuring a related construct. Items were extracted regardless of whether the instrument had been validated, as many constructs in human-AI interaction are still assessed with purpose-built scales. The resulting initial item pool contained 1429 items. 

\paragraph{Generalizing and Unifying Item Phrasing.}
The items were then rephrased as generalized statements: specific tasks (e.g., writing, driving) and domains were replaced with \texttt{this task}, deliverables (e.g., designs) with \textit{outcome}, and human roles (e.g., doctor, colleague) with \textit{human professional} or \textit{human}. Named systems (e.g., ChatGPT) and technology labels *e.g., AI) were replaced with a \texttt{[this system/agent]} placeholder to be adjusted for the specific system or agent under evaluation. Items that could not be generalized to interactions in a different user study were removed. Given the size of the pool, one author rephrased 60
representative items by hand. Using these as a reference, an LLM (Claude Sonnet
5) rephrased 120 additional items, which the author then edited. Based on this, we
tuned the instructions for the LLM and used the 200 resulting items as a 
reference for rephrasing the remainder automatically.

\paragraph{Removal of Redundant and Irrelevant Items.} After rephrasing, repetitive items became evident. Four authors worked together to remove theses and items clearly unrelated to perceived value alignment ($n=1045$). From the remaining 384 items, two authors removed items still
presupposing specific interaction modalities (e.g., chat) or use cases, as these limit the generalizability of the scale ($n=110$). The 274 remaining items were checked and, if needed, edited for correctness
and naturalness by a native English speaker among the authors and constituted
the initial pool.

\section{Content Validity}

We assessed its content validity in an expert study before testing the item pool with users. Following recommendations by ~\citet{boateng_best_2018} we recruited domain experts to judged whether the items capture the construct.

\subsection{Participants}

Our inclusion criterion was demonstrated domain expertise, operationalized as having authored at least one peer-reviewed publication in value alignment or in (psychometric) scale development. We recruited experts through our professional network. Participation was voluntary and uncompensated. Seven experts completed the survey (3 men, 4 women; $M_{age} = 32.7$, $SD = 6.4$, range 25--41). On a 5-point agreement scale, four participants rated themselves as experts in scale development and four in AI alignment. \autoref{tab:expert-demographics} reports individual-level demographic and self-rated expertise data.

\begin{table}[htb]
\centering
\caption{Expert sample for the content-validity assessment ($N = 7$). Scale-development (Scale dev.) and AI-alignment (AI align.) expertise were self-rated on a 5-point agreement scale (1 = strongly disagree, 5 = strongly agree).}
\label{tab:expert-demographics}
\setlength{\tabcolsep}{2pt}
\begin{tabular*}{\linewidth}{@{\extracolsep{\fill}}lccllcc@{}}
\toprule
   &        &     &          &        & Scale & AI     \\
ID & Gender & Age & Position & Degree & dev.  & align. \\
\midrule
P1 & M & 41 & Faculty        & Doctoral & 4 & 4 \\
P2 & M & 40 & PhD researcher & Master's & 5 & 4 \\
P3 & F & 27 & PhD researcher & Master's & 1 & 4 \\
P4 & F & 28 & Postdoc        & Doctoral & 2 & 3 \\
P5 & M & 36 & Postdoc        & Doctoral & 1 & 3 \\
P6 & F & 25 & PhD researcher & Master's & 4 & 5 \\
P7 & F & 32 & PhD researcher & Master's & 4 & 3 \\
\bottomrule
\end{tabular*}
\end{table}

\subsection{Procedure}

Experts first provided informed consent and reported demographic information (gender, age, highest degree, current position) and their self-rated expertise in scale development and AI alignment on a 5-point scale.

Experts read the construct definition of perceived value alignment and then rated each of the 274 candidate items. For every item, experts saw the generalized item statement (e.g., ``[This system/agent] understands my needs.'') and answered ``How relevant is the statement above for measuring perceived value alignment?'' on a 4-point scale ranging from 1 (not relevant) to 4 (highly relevant), with an optional free-text field to comment on that item. After rating the full pool, experts had the option to answer whether they felt any factors of perceived value alignment were missing from the item pool. Finally, they were presented with two example scenarios for scale administration and asked to comment what other use cases, interactions, and scenarios might be important to cover.

\subsection{Data Analysis}

For each item, we computed the mean relevance rating across the seven experts. On our 1--4 scale, a mean above the scale midpoint (2.5) indicates that, in aggregate, experts judged the item more relevant than irrelevant to perceived value alignment. Items below the midpoint were dropped. This retained 106 of the 274 items (38.7\%) for the item-testing study.

We analyzed the expert suggestions qualitatively and found a number of specific tasks and domains, from health care to design.
In addition, specific types of values emerged (ethical, instrumental preferences, moral, social, epistemic, privacy, and autonomy). One expert argued for varying scenarios along dimensions such as embodiment, from screen-based software to physically embodied agents, the distribution and level of agency, from responding with recommendations to full autonomy, duration of the relation, from one-shot to spanning a longer time, and stakeholder structure, from one user to balancing multiple user groups.

\section{Item Testing}
\label{sec:item-testing}

\subsection{Participants}

We recruited 673 English-speaking participants from the USA and the UK via the recruiting platform Prolific\footnote{https://www.prolific.com/, accessed on Sep 3, 2026}, of whom 601 passed the comprehension check and attention checks and completed the full questionnaire. The participants were compensated with \pounds 2.25 and took a median time of 12 minutes to complete the survey. 

\subsection{Materials}
We selected six scenarios aiming to span the space suggested during the expert study. An \textbf{autonomous car} and a \textbf{household robot} represented physically embodied, high-autonomy agents, contrasted with a screen-based \textbf{Photo editing interaction} and a \textbf{Chat-based travel agent}, which vary in agency from a generative editing assistant to a recommendation-only itinerary planner. A \textbf{Medical care robot} and \textbf{Harmful conversational agent} extend the value range beyond task performance into ethical conflict, varied stakeholder structure, and misalignment: the former enforces health goals defined by a purchaser against the end user's will, the latter voices misaligned ideas derived from \citet{betley_emergent_2026}. The car, robot, and photo-editing interactions were presented as videos, and the chat interactions were graphical layouts with text. These materials were shown to the participants to evoke a perception of value alignment. Full stimulus descriptions, materials, and generation details are reported in the supplementary materials.

\subsection{Procedure}
After giving informed consent, participants completed a task comprehension test
verifying that they engaged with the scenarios and comprehended the
instructions: they watched a short video of a human--AI interaction and
answered a question about it. Participants who passed within two attempts
were randomly assigned one of the scenarios and answered two questions about
its content to ensure they had watched or read it. Correct answers were required to proceed but following Prolific guidelines, mistakes did not result in exclusion at this later stage of the survey. After observing and understanding the scenario, participants rated their agreement with each of the 106 items on a five-point Likert scale.

\section{Item Reduction}
\label{sec:item-reduction}

We analyzed the data collected in the previous step to identify the latent structure underlying the item pool and reduce it to a manageable scale. This was done using Exploratory Factor Analysis (EFA), which models the observed items as manifestations of a smaller number of latent factors. 

\subsection{Data Screening}

Since careless responding is a prevalent issue in survey data~\cite{ward_dealing_2023, belliveau_evaluating_2022}, we screened the data for failed attention checks and implausibly fast responses averaging less than 2 seconds per item. In addition, we computed longstrings: the longest streak of consecutive identical answers in each response. Based on the number of negative items and a plot of longstrings in the sample, we identified 20 as the threshold that careful item-based responses would not exceed. In total, 88 responses (14.5\%) were removed, leaving 518 respondent-scenario rows for analysis.

\subsection{Item Pre-processing and Adequacy Testing}

Before conducting the EFA, we reverse-coded all negatively worded items so that higher scores consistently reflected the same underlying direction across the pool. We then inspected each item's mean, skew, kurtosis, and corrected item-total correlation, removing three items whose corrected item-total correlation fell below .30 ($r_{drop} < .30$), indicating poor discrimination relative to the rest of the pool. This left 103 items for factor analysis.

We then evaluated the suitability of the retained item set for factor analysis using the Kaiser-Meyer-Olkin (KMO) measure of sampling adequacy, which compares each item's raw inter-item correlations to its partial correlations after controlling for all other items (values closer to 1.0 indicate the item shares variance directly with the pool rather than only through intermediaries). The overall KMO Measure of Sampling Adequacy was $MSA = .985$, well above the conventional .60 threshold. We complemented this with Bartlett's Test of Sphericity, which tests whether the correlation matrix differs significantly from an identity matrix. The test was significant, $\chi^2(5253) = 49706.5$, $p < .001$, confirming that the items were sufficiently intercorrelated to support factor extraction.

\begin{figure}
    \centering
    \includegraphics[width=\linewidth]{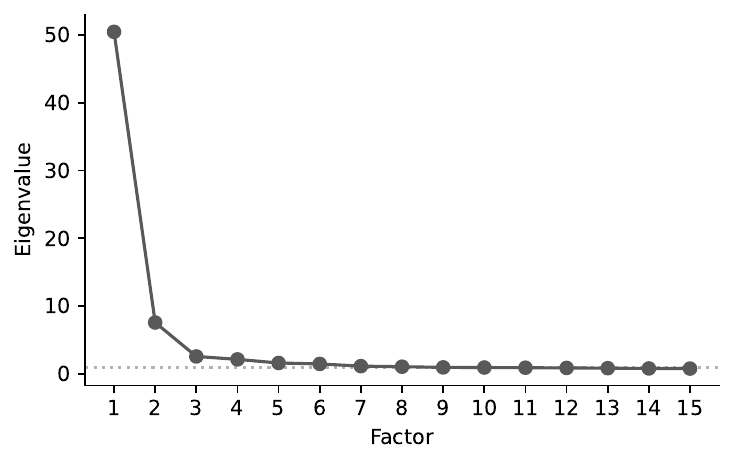}
    \caption{Scree plot showing eigenvalues near one for the number of factors over two, indicating a 2-factor structure. }
    \label{fig:scree}
    \Description{Line plot of a scree test showing eigenvalues on the vertical axis against factor numbers 1 through 15 on the horizontal axis. The curve drops sharply from an eigenvalue over 50 for the first factor to around 7.5 for the second factor, followed by an elbow that levels off below 3 for factors 3 through 15, indicating a distinct two-factor structure.}
\end{figure}

\subsection{Exploratory Factor Analysis}
\begin{table*}
\centering
\caption{The 15-item, two-factor \scaleshort{} solution from the EFA: \factora{} (\fa{}) and \factorb{} (\fb{}). Secondary loadings are shown in gray. Primary loadings used for scoring are in \textbf{bold}.}
\label{tab:Factors}
\setlength{\tabcolsep}{4pt}
\begin{tabularx}{\linewidth}{@{}l>{\raggedright\arraybackslash\hangindent=1em\hangafter=1}Xrr@{}}
\toprule
\textbf{Code} & \textbf{Item} & \textbf{\fa{}} & \textbf{\fb{}} \\
\midrule
$VM_1$ & [This system/agent] prefers a different outcome than I do in this context. & \textcolor{gray}{$-$.06} & \textbf{.82} \\
$VM_2$ & I was disappointed by how [this system/agent] interpreted my input. & \textcolor{gray}{.23} & \textbf{.77} \\
$VM_3$ & I reject [this system/agent]'s preferred action. & \textcolor{gray}{.19} & \textbf{.77} \\
$VM_4$ & [This system/agent]'s preferred outcomes and mine are in conflict in this context. & \textcolor{gray}{$-$.06} & \textbf{.77} \\
$VM_5$ & [This system/agent] led me away from what I wanted. & \textcolor{gray}{$-$.11} & \textbf{.75} \\
$VM_6$ & The outcome mismatched what I had aimed for in this task. & \textcolor{gray}{.10} & \textbf{.72} \\
$VM_7$ & I can influence [this system/agent]'s decision-making process and outcomes. & \textcolor{gray}{.18} & \textbf{.53} \\
\addlinespace[1em]
$VU_1$ & [This system/agent] understands how its actions affect me. & \textbf{.79} & \textcolor{gray}{.09} \\
$VU_2$ & [This system/agent] is aware of the meaning in our interaction. & \textbf{.77} & \textcolor{gray}{.05} \\
$VU_3$ & [This system/agent] understands things as a human would. & \textbf{.77} & \textcolor{gray}{.03} \\
$VU_4$ & [This system/agent] can put itself in my shoes. & \textbf{.74} & \textcolor{gray}{.08} \\
$VU_5$ & [This system/agent] understands the world around us. & \textbf{.71} & \textcolor{gray}{.11} \\
$VU_6$ & [This system/agent] and I have similar identities. & \textbf{.70} & \textcolor{gray}{.07} \\
$VU_7$ & [This system/agent] knows why I prefer a certain action. & \textbf{.69} & \textcolor{gray}{.16} \\
$VU_8$ & [This system/agent] was able to realize that we had different approaches to achieving the goal. & \textbf{.62} & \textcolor{gray}{$-$.05} \\
\bottomrule
\end{tabularx}
\end{table*}

To determine how many factors to retain, we compared two quantitative criteria (the Kaiser criterion (eigenvalue $>$ 1), and parallel analysis) against visual inspection of the scree plot. The two criteria disagreed, suggesting 8 (Kaiser), or 6 (parallel analysis). With a large, semantically overlapping item pool, this is expected, since all three remain sensitive to numerous small, secondary factors that do not represent a stable or interpretable structure \cite{Horn1965ARA, Cattell1977ACT}. The scree plot, by contrast, showed an elbow after the second factor: eigenvalues fell from 7.58 (Factor 2) to 2.57 (Factor 3) before leveling off. Given this inflection, we retained a two-factor solution.

Factors were extracted using minimum residual (minres) estimation. For the rotation, we applied varimax rather than an oblique method. Varimax constrains the resulting factors to be uncorrelated, which is a stronger assumption than the data may fully satisfy. We selected varimax because it yielded a more balanced distribution of retained items across these facets. Item selection prioritized strong primary loadings and low cross-loadings. Orthogonality was treated as a simplifying assumption, acknowledging that the two perceptions may still correlate.

Based on the rotated solution, we retained items with a factor loading of at least .60, a cross-loading no higher than .20, and a gap between these two of at least .30. Of the 103 candidate items, 32 were dropped for an insufficient factor loading and 56 for cross-loading, leaving a final scale of \textbf{15 items across two factors}, with 8 and 7 items, respectively. Re-fitting the model on this final item set, the two-factor solution accounted for 55.0\% of the common variance (29.2\% and 25.8\% per factor), with communalities ranging from $h^2 = .32$ to $.67$. \autoref{tab:Factors} reports the resulting pattern matrix.

We interpreted the content of each factor's items. The first factor with 8 items described whether the system or agent empathizes with the user (``put itself in my shoes'') and ``understands'' the human or context to the level of reasons (``knows why I prefer a certain action,'' ``understands how its actions affect me''), shared perspective (``understands things as a human would''), and relational context (``aware of the meaning in our interaction''), leading us to name this factor \factora{} (\fa{}). The second factor, with 7 items, describes whether the system or agent aligns with user intent (``prefers a different outcome,'' ``in conflict''), successfully executes intended goals (``disappointed,'' ``led me away,'' ``mismatched''), and incorporates user input (``influence'') in practice, leading us to name this factor \factorb{} (\fb{}).

\section{Scale Evaluation}
\label{sec:evaluation}

To evaluate the factor structure of the \scaleshort{}, we conducted confirmatory factor analysis (CFA). Further, we evaluated the internal consistency of the two factors and compared \scaleshort{} to scales on related constructs to assess discriminant and convergent validity. We collected an independent sample and altered the set of scenarios to include two pairs of use cases, manipulating the level of alignment for one or both factors to support a known-group validation. We did not include a re-test in the evaluation, as perceived value alignment is influenced by interactions rather than representing a stable characteristic of participants. Finally, we used a scale abbreviation method to develop \scaleshort{} short and re-ran internal consistency, discriminant, convergent, and known-group validity checks to ensure it maintained comparable performance.

\subsection{Participants}
We recruited 357 English-speaking participants from the USA and the UK via Prolific\footnote{https://www.prolific.com/, accessed on Sep 3, 2026}, of whom 296 passed the comprehension check and completed the full questionnaire. The participants were compensated with \pounds 2.25 and took a median time of 9 minutes to complete the survey. 

\subsection{Materials}
The scenarios comprised two complementary pairs (four groups in total) for known-group comparison: a more- and a less-aligned version of an autonomous car and a chat-based AI travel agent. Controlling other variables, we picked two instances of the autonomous car performing an unprotected left turn, filmed from inside the car with a hand-held camera. In one condition, the car performs a risky maneuver, while in the other, it carefully waits for oncoming traffic to stop before driving. The AI travel agent was prompted with the same details for a ``budget trip'' to ``London'' but in one case mistakenly identifies the desired destination as London, Ontario, and ignores the user's preference for economical pricing. 

\subsection{Measures}

To test the discriminant and convergent validity of the \scaleshort{}, we included three related instruments from the literature, with their subscales:

\begin{itemize}
    \item[\textbf{TiA}] Trust in Automation~\cite{körber_theoretical_2019}: a measure of trust in an automated system with the six factors \textbf{Reliability/Competence}, \textbf{Understanding/Predictability}, \textbf{Familiarity}, \textbf{Intention of Developers}, \textbf{Propensity to Trust}, and \textbf{Trust in Automation}. The scale was included to evaluate convergent and discriminant validity, as trust is expected to correlate positively with perceived alignment (convergent validity) despite representing distinct constructs across the subscales (discriminant validity). The strongest correlation is expected between the TiA-Reliability/Competence factor and \fb, as both are concerned with the system producing desired outcomes.
    \item[\textbf{PC}] \enlargethispage{1\baselineskip}
    \parbox[t]{\linewidth}{Perceived Creepiness of Technology~\cite{woźniak_creepy_2021}: a three-factor measure of the unease a technology evokes. Consisting of \textbf{Implied Malice}, \textbf{Undesirability}, and \textbf{Unpredictability}. The PC was included for convergent validity, as a system misaligned with users is likely to cause unease and thus negatively correlate with PC.}
    \item[\textbf{ATI}] Affinity for Technology Interaction~\cite{franke_personal_2019}: a one-factor measure of a person's general disposition to engage with technology. This scale was included to assess discriminant validity, as affinity to technology is a general disposition that should be more stable than perceived alignment. 
\end{itemize}

\subsection{Procedure}
Participants provided informed consent and completed the task comprehension test from the item testing study (\ref{sec:item-testing}). Those who passed were randomly assigned to one of the four groups, viewed the scenario video or text, and answered two attention check questions about the content before completing the scales. To prevent order effects, the four scales (TiA, PC, ATI, and \scaleshort{}) were administered in random order.

\subsection{Data Screening}
Following our EFA data screening procedures (\ref{sec:item-reduction}), we removed participants showing signs of careless responding based on attention checks, completion times, and longstrings per scale. We excluded participants who failed one attention check, averaged under 2 seconds per item, or produced longstrings with a probability under 0.1 given the proportion of reverse-coded items. Screening only the \scaleshort{} scale resulted in removing 41 samples for the CFA, whereas multi-scale validation required removing 75 samples that failed criteria on any evaluated scale. The sample used for validation consisted of 64 in the travel agent England scenario, 55 in the travel agent Ontario scenario, 55 in the risky car scenario, and 51 in the careful car scenario.

\subsection{Confirmatory Factor Analysis}
\begin{table*}[htb]
\centering
\caption{Standardized factor loadings for the final 13-item, two-factor confirmatory model. $\diamond$ reverse coded}
\label{tab:CFA}
\renewcommand{\arraystretch}{1.3}
\begin{tabularx}{\linewidth}{Xcccc}
\toprule
Item & Code & Std.\ loading & 95\% CI & Source \\
\midrule
{}[This system/agent] prefers a different outcome than I do in this context.$\diamond$ & $VM_1$ & \textbf{.69} & [.59, .79] & \cite{woide_measuring_2021} \\
I was disappointed by how [this system/agent] interpreted my input.$\diamond$ & $VM_2$ & \textbf{.84} & [.77, .91] & \cite{hegemann_tomigo_2026} \\
I reject [this system/agent]'s preferred action.$\diamond$ & $VM_3$ & \textbf{.84} & [.77, .92] & \cite{woide_measuring_2021} \\
{}[This system/agent] led me away from what I wanted.$\diamond$ & $VM_5$ & \textbf{.80} & [.73, .88] & \cite{hegemann_tomigo_2026} \\
The outcome mismatched what I had aimed for in this task.$\diamond$ & $VM_6$ & \textbf{.83} & [.77, .90] & \cite{hegemann_tomigo_2026}\\
\midrule
{}[This system/agent] understands how its actions affect me. & $VU_1$ & \textbf{.79} & [.73, .85] & \cite{woide_measuring_2021}\\
{}[This system/agent] is aware of the meaning in our interaction. & $VU_2$ & \textbf{.71} & [.63, .80] & \cite{liang_perceived_2025}\\
{}[This system/agent] understands things as a human would. & $VU_3$ & \textbf{.77} & [.70, .83] & \cite{liang_perceived_2025}\\
{}[This system/agent] can put itself in my shoes. & $VU_4$ & \textbf{.78} & [.72, .85] & \cite{zengzhan_impact_2026} \\
{}[This system/agent] understands the world around us. & $VU_5$ & \textbf{.73} & [.66, .80] & \cite{liang_perceived_2025}\\
{}[This system/agent] and I have similar identities. & $VU_6$ & \textbf{.67} & [.60, .75] & \cite{yang_who_2026} \\
{}[This system/agent] knows why I prefer a certain action. & $VU_7$ & \textbf{.70} & [.61, .80] & \cite{flathmann_teammate_2026}\\
{}[This system/agent] was able to realize that we had different approaches to achieving the goal. & $VU_8$ & \textbf{.47} & [.34, .59] & \cite{attig_measuring_2026} \\
\bottomrule
\end{tabularx}
\end{table*}

The correlated two-factor model's largest modification index flagged items $VM_4$ and $VM_1$ (MI = 19.7), indicating item redundancy. Rather than freeing this residual covariance, we removed the weaker-loading item of the pair, item 4 (EFA loading .70 vs. item 29's .74). In the resulting 14-item model, item $VM_7$ loaded at only .27 and was removed as items scoring $<0.3$ are considered inadequate~\cite{boateng_best_2018}.

The final model, with the two items removed, comprised 13 items across the same two correlated factors (\fa{}: 8 items; \fb{}: 5 items) and showed a good fit: $\chi^2(64) = 114.7$, CFI = .969, TLI = .962, RMSEA = .058, SRMR = .047 (MLR). The WLSMV robustness check agreed on CFI, TLI, and SRMR (.982, .978, .057) and came close on RMSEA (.074). \autoref{tab:CFA} reports the standardized loadings and confidence intervals for this final model.

\subsection{Internal Consistency}
\label{sec:internalcons}

We assessed reliability for both factors of the long-form (13-item) alignment scale using Cronbach's $\alpha$ with a 95\% CI, and complemented it with composite reliability ($\omega$) and average variance extracted (AVE) from each factor's own single-factor CFA, since $\alpha$ assumes equal item loadings and our loadings are not equal \citep{boateng_best_2018}. Both factors exceeded the .70 acceptable threshold by a wide margin, and neither approached the .95 redundancy warning: \fb{} ($\alpha = .905$) and \fa{} ($\alpha = .890$). \autoref{tab:reliability} reports the full set of estimates.
\begin{table}[htb]
\centering
\caption{Internal consistency of the long-form (13-item) \scaleshort{} factors.}
\label{tab:reliability}
\begin{tabular}{lccccc}
\toprule
Factor & $k$ & $\alpha$ & 95\% CI & $\omega$ & $AVE$ \\
\midrule
\fb{} & 5 & \textbf{.905} & [.88, .92] & .906 & .658 \\
\fa{} & 8 & \textbf{.890} & [.87, .91] & .891 & .510 \\
\bottomrule
\end{tabular}
\end{table}

\subsection{Discriminant Validity}
\label{sec:discriminantval}

To establish discriminant validity, both alignment factors were evaluated against Perceived Creepiness (PC) and Affinity for Technology Interaction (ATI). Because ATI captures a stable general disposition and PC measures specific technological unease, both were expected to diverge from our situational alignment constructs. Both alignment factors remained comfortably below the strict .85 discriminant validity threshold~\cite{henseler_new_2015} across all subscales of PC (.425–.579) and ATI (.150–.188). The \textit{Familiarity}, \textit{Intention of Developers}, \textit{Propensity to Trust}, and \textit{Trust in Automation}, \textit{Understanding/Predictability} also stayed comfortably below the strict threshold (HTMT = .236–.707), further establishing discriminant validity. \textit{Understanding/Predictability} correlated more strongly with \fb{} than the other dimensions, as expected. \autoref{tab:discriminant} reports the complete HTMT matrix. \autoref{tab:discriminant} reports the complete HTMT matrix. 

\begin{table}
\centering
\caption{HTMT between the \scaleshort{} factors and each comparison-instrument subscale (\textit{N} = 225).}
\label{tab:discriminant}
\begin{tabularx}{\linewidth}{@{}Xcc@{}}
\toprule
Instrument and subscale & \fb{} & \fa{} \\
\midrule
\textit{Trust in Automation (TiA)} & & \\
\quad Reliability/Competence         & .759 & .640 \\
\quad Understanding/Predictability   & .707 & .503 \\
\quad Familiarity                    & .236 & .345 \\
\quad Intention of Developers        & .475 & .577 \\
\quad Propensity to Trust            & .518 & .578 \\
\quad Trust in Automation (subscale) & .627 & .649 \\
\addlinespace
\textit{Perceived Creepiness (PC)} & & \\
\quad Implied Malice                 & .507 & .425 \\
\quad Undesirability                 & .579 & .477 \\
\quad Unpredictability               & .546 & .429 \\
\addlinespace
\textit{Affinity for Technology Interaction (ATI)} & & \\
\quad Single factor                  & .188 & .150 \\
\bottomrule
\end{tabularx}
\end{table}

\subsection{Convergent Validity}
\label{sec:convergentval}

We predicted that \fb{} would correlate most strongly with TiA's \textit{Reliability/Competence} subscale, as both relate to producing desired outcomes, and that perceived alignment would correlate positively with trust (TiA) and negatively with PC and show little or no correlation with ATI. Overall, perceived alignment demonstrated the expected pattern of convergent and discriminant relationships. Both factors correlated positively with total trust (\fb{}: $r = .637$; \fa{}: $r = .595$) and negatively with total creepiness (\fb{}: $r = -.538$; \fa{}: $r = -.449$). Across all measures, \fb{} consistently exhibited stronger associations with trust and creepiness dimensions than \fa{}. Finally, correlations with general technology affinity (ATI) were small or non-significant (\fb{}: $r = .169$; \fa{}: $r = .111, 95\%\text{ CI } [-.02, .24]$), supporting the discriminant validity of both alignment factors against general user dispositions. \fb{} correlated most strongly with Trust-Reliability/Competence ($r = .671$, $p < .001$) among all Trust subscales. 

\begin{table}[htb]
\centering
\small
\caption{Convergent and discriminant correlations for the long-form \scaleshort{} factors (\textit{N} = 225).}
\label{tab:convergent}
\setlength{\tabcolsep}{3pt}
\begin{tabularx}{\linewidth}{@{}>{\raggedright\arraybackslash\hangindent=2em\hangafter=1}Xrcrc@{}}
\toprule
 & \multicolumn{2}{c}{\fb{}} & \multicolumn{2}{c@{}}{\fa{}} \\
\cmidrule(lr){2-3}\cmidrule(l){4-5}
Comparison scale & \multicolumn{1}{c}{$r$} & 95\% CI & \multicolumn{1}{c}{$r$} & 95\% CI \\
\midrule
\textit{Trust in Automation (TiA)} & & & & \\
\quad Reliability/Competence       & .671 & [.59, .74] & .570 & [.48, .65] \\
\quad Understanding/ Predictability & .607 & [.52, .69] & .438 & [.33, .54] \\
\quad Total                        & .637 & [.55, .71] & .595 & [.50, .67] \\
\addlinespace
Perceived Creepiness, total        & $-$.538 & [$-$.62, $-$.44] & $-$.449 & [$-$.55, $-$.34] \\
\addlinespace
Affinity (ATI)                     & .169 & [.04, .29] & .111 & [$-$.02, .24] \\
\bottomrule
\end{tabularx}
\end{table}

\subsection{Known Groups Validity}
\label{sec:knowngroupsval}
\begin{figure*}
    \centering
    \includegraphics[width=0.88\linewidth]{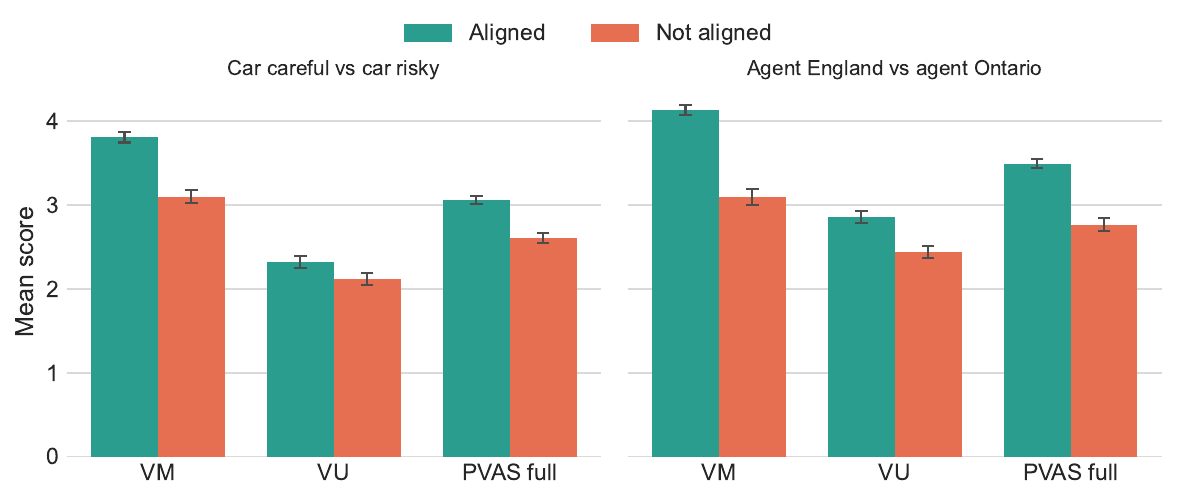}
    \caption{The mean \fb, \fa, and combined scores of the careful car and travel agent recommending sights in London, England (aligned) compared to their misaligned controls (car risky, agent Ontario).  }
    \label{fig:knowgroup}
    \Description{Bar charts displaying the mean scores for Value Manifestation (VM), Value Understanding (VU), and overall PVAS across two scenario comparisons, with error bars representing confidence intervals. The left plot compares the careful car in green against the risky car in orange, showing significantly higher VM, slightly higher VU, and higher overall PVAS for the careful car. The right plot compares the travel agent specifying London, England in green against London, Ontario in orange, showing substantially higher VM, higher VU, and higher overall PVAS for the aligned England agent.}
\end{figure*}

To ensure adequate statistical power to detect moderate group differences, an additional sample ($N = 495$) was collected specifically for this comparison. The protocol remained identical to that of the primary study, except that participants completed only the PVAS. This reduced the average completion time to 5 minutes, and participants received \pounds0.75 in compensation. Applying the same data screening procedures yielded a combined final sample of $N = 658$ for the known-groups comparison.

We tested two contrasting scenarios: \textit{car risky} vs. \textit{car careful} (both performing a left turn but handling the interaction with traffic differently) and \textit{agent england} vs. \textit{agent ontario} (both planning an itinerary and accommodation in London but one misunderstanding the state). We used Welch's \textit{t}-test with Cohen's \textit{d} and a 95\% CI. 

\fb{} showed a large effect in both contrasts (car: $d = .806$, 95\% CI [.58, 1.04], $p < .001$ and agent: $d = .956$, 95\% CI [.73, 1.18], $p < .001$), consistent with the vignettes directly manipulating value manifestation through observable behavior. \fa{} also showed significant effects, which were smaller in the car scenario ($d = .238$, 95\% CI [.02, .46], $p = .035$) and a medium in the travel agent scenario ($d = .453$, 95\% CI [.24, .67], $p < .001$). All details can be seen in Table \ref{tab:knowngroups}.

\begin{table}[htb]
\centering
\caption{Known-groups contrasts for \scaleshort{} on the known-group samples (\textit{N} = 658).}
\label{tab:knowngroups}
\begin{tabularx}{\linewidth}{@{}Xccr@{}}
\toprule
Contrast and measure & $d$ & 95\% CI & \multicolumn{1}{c@{}}{$p$} \\
\midrule
\textit{Car risky} vs.\ \textit{Car careful} & & & \\
\quad \fb{}               & .806 & [.58, 1.04] & $<.001$ \\
\quad \fa{}               & .238 & [.02, .46]  & .035    \\
\quad \scaleshort{} total & .644 & [.42, .87]  & $<.001$ \\
\addlinespace
\textit{Agent England} vs.\ \textit{Agent Ontario} & & & \\
\quad \fb{}               & .956 & [.73, 1.18] & $<.001$ \\
\quad \fa{}               & .453 & [.24, .67]  & $<.001$ \\
\quad \scaleshort{} total & .833 & [.61, 1.05] & $<.001$ \\
\bottomrule
\end{tabularx}
\end{table}

\newpage
\subsection{Short Form}
\label{sec:short-form}

To reduce participant burden, we shortened the 13-item long form using the genetic algorithm method of \citet{yarkoni_abbreviation_2010} on the combined EFA and CFA samples (\textit{N} = 777). For each factor, we searched for the item subset that best reproduced the full-factor score under a loss function balancing model fit ($R^2$) against a fixed item cost of 0.05 following ~\cite{yarkoni_abbreviation_2010}. The inclusion search plateaued at three items per factor: \fb{} ($R^2 = .941$, items $VM_1$, $VM_3$, $VM_5$) and \fa{} ($R^2 = .894$, $VU_1$, $VU_5$, $VU_7$), yielding a 6-item short form.

To verify that item reduction did not compromise the psychometric properties of \scaleshort{}, we re-assessed the validation metrics with the same tests and samples as described in Subsections \ref{sec:internalcons}--\ref{sec:knowngroupsval}. The short-form factor scores correlated strongly with their long-form counterparts ($r = .975$ for \fb{}, $r = .951$ for \fa{}, $r = .960$ total), indicating low information loss. Internal consistency attenuated as expected with fewer items, but both subscales retained adequate reliability (\fb{}: $\alpha = .843$, $\omega=0.844$; \fa{}: $\alpha = .794$, $\omega=0.796$). Convergent validity remained robust, with the short form displaying strong positive correlations with overall trust ($r = .73$) and its subscales, negative correlations with perceived creepiness ($r = -.60$), and low association with general technology affinity ($r = .16$). Discriminant validity was similarly supported, as HTMT ratios between short-form factors and external subscales remained below the conservative threshold of $0.85$—ranging from $0.138$ to $0.772$ for the included subscales. Finally, known-groups validation confirmed that the short form maintained sensitivity to deliberate variations in system alignment: total scale scores differentiated between aligned and misaligned conditions in both the driving scenario ($d = 0.618, p < .001$) and the travel agent scenario ($d = 0.807, p < .001$). Subscale sensitivity was mostly preserved: \fb{} scores showed large, significant differences in both the driving ($d = 0.818, p < .001$) and travel agent ($d = 0.967, p < .001$) scenarios, while \fa scores captured moderate differences in the travel agent scenario ($d = 0.417, p < .001$). The difference in the car scenario became slightly smaller and was not significant ($d = 0.200, p = .076$).

\section{Discussion}
In this paper, we present the development and validation of a 13-item scale to measure value alignment in HAI. Through factor analysis, we identified a two-factor structure: VM, which measures the perceived manifestation of the system's values, and VU, which measures the extent to which the system ``\textit{understands}'' the user, that is, recognizes, interprets, and correlates with the user's values and intentions. 
We abbreviated the scale further to a 6-item short-form. 

\subsection{Using the Scale in Training AI Systems}
How do we train AI systems to reflect what their users value? Today's training pipelines answer this question with stand-ins. A system is rewarded for what raters prefer, for what a grader marks as correct, or for what satisfies a rubric someone wrote for the task. The user is not asked whether the system understood what they care about or acted on it. The \scaleshort{} is a validated way to ask, and future work should test whether it can be used as a training signal.

We see two ways to do this. In the first, a grading model reads a transcript of the interaction and answers the items on the user's behalf, and the scores are used as the reward. Whether a grader's answers match the user's own is unknown; this can be tested by collecting user responses on the same interactions and comparing them. In the second, users complete the six-item short form after an interaction, and a reward model is trained to predict their responses instead of pairwise preferences. This is more expensive per label but measures the construct directly. In both cases the scale measures perceived alignment, and a system trained to raise its score may learn to appear understanding rather than to understand. A signal built from the scale should be paired with behavioral measures that do not depend on the user's impression, and the validity of the scale should be re-examined once systems have been optimized against it.

\subsection{Scale Scoring and Interpretation of Factors}
\label{sec:scoring}
All PVAS items are answered on a five-point Likert scale ranging from Strongly Disagree (1) to Strongly Agree (5).
The five VM items ($VM_1$, $VM_2$, $VM_3$, $VM_5$, $VM_6$) are negatively worded and are reverse-scored as $6 - x$ before any score is computed.
The eight VU items are scored as recorded.
After reversal, higher values consistently indicate that a user perceives greater value alignment.

Each factor score is the arithmetic mean of its items:
\begin{align*}
    VU &= \frac{1}{8}\sum_{i=1}^{8} VU_i, &
    VM &= \frac{1}{5}\sum_{i=1}^{5} VM_{i_R}
\end{align*}

The overall score is the mean of the two factor scores:
\begin{align*}
    PVAS &= \frac{VU + VM}{2}
\end{align*}

We average the factor scores rather than the 13 items, so that both components contribute equally despite the unequal number of items per factor.
The overall score expresses the degree to which a user perceives a system as aligned with their values.
High scores indicate that users experienced the system as recognizing what matters to them and as acting accordingly; low scores indicate that they did not.
Since no normative data exist, scores are interpreted comparatively, between systems, conditions, or points in time, rather than against an absolute criterion.

The six-item short form is scored identically, with three items per factor.
Its scores track the long form closely at the person level ($r = .975$ for VM, $r = .951$ for VU, and $r = .960$ for the overall score), so the interpretation above applies unchanged.

We recommend reporting the two factor scores alongside the overall score.
The overall score shows how aligned a system is perceived to be, but analyzing the factors individually might shed light on how to improve an interactive AI system. 
A low VM combined with a high VU could indicate that the system understands the user but is not capable of performing the required actions.
In such a case, understanding might be communicated, and the user might perceive it as this, while its manifestation in behavior diverges, for example, because the system lacks the capability or the permissions to act accordingly.
A low VU combined with a high VM indicates the reverse: the system behaves acceptably but does not convey that it recognized what mattered, which concerns how it communicates, explains, and elicits.
In both cases, the possibility to differentiate the two factors can guide further evaluation and system design.

\subsection{Factors and Relation to Related Work}
\label{sec:factors-relatedwork}
Our construct definition rests on a three-level hierarchy of values, goals, and preferences, but the factor analysis did not reproduce these levels.
It separated whether a system apprehends what matters to its user (VU) from whether its actions reflect it (VM).
The mapping is asymmetric.
All VM items are situated, referring to outcomes, actions, and preferences in a given context, which follows from the construct stating that behavior can only be judged at the level of goals and preferences.
The VU items address the level above, asking whether the system grasps what matters and why concrete choices are made.
This is consistent with our definition, in which values become observable only through their expression in context, meaning that users cannot judge value congruence directly but need to infer it.

\subsection{Implications}
As agents gain autonomy, they take over trade-offs that users previously made themselves. PVAS allows practitioners to compare how users experience different ways of delegating those decisions: an agent might infer priorities from past behavior, ask when priorities conflict, or let users revise its choices. The same assessment can compare models before and after fine-tuning, examining whether adaptation improves VU, VM, or both. Reporting these separately preserves a distinction that matters for implementation: users may feel understood while rejecting what the agent does.

For researchers, PVAS enables experiments that distinguish the effects of how an agent communicates from those of what it does. Explanations might increase VU without improving VM, while decisions that reflect users' priorities might improve VM without strengthening the feeling of being understood. PVAS could also support the development of personalized models, such as digital twins intended to reflect an individual's values, goals, and preferences. Ratings from that individual would provide an empirical basis for assessing whether the model is perceived as understanding and acting on what matters to them. These assessments could guide the evaluation of individual personalization and help determine whether successive model versions improve perceived alignment with the person they represent.

\section{Limitations and Future Work}

Several limitations of this work point to opportunities for future research.

Our evaluation relied primarily on scenario-based vignettes such as videos and text rather interaction with functional systems. While vignettes allowed precise experimental control across modalities and alignment levels, perceived alignment in dynamic, multi-turn applications may introduce behavioral factors not fully captured in the vignette setups. Future work should validate the scale in naturalistic, interactive deployments.  

Our sample was limited to English-speaking participants from the US and the UK. Because personal values and expectations of agent behavior may vary across cultures, further research could validate PVAS across more cultural groups.  

Although the 6-item short form was derived using algorithmic optimization and demonstrated high statistical correspondence with the full scale, it was evaluated on the same datasets used for its construction. Independent empirical validation of the short form in standalone user studies would further increase confidence in the short form of \scaleshort{}.

\section{Conclusion}

We introduced the Perceived Value Alignment Scale (\scaleshort{} ), a psychometrically validated 13-item instrument, and its 6-item short form, to measure perceived value alignment in human-AI interaction across two distinct dimensions: Value Understanding and Value Manifestation. By offering a reliable, domain-general metric grounded in observed agent behavior, \scaleshort{} enables researchers to evaluate interactive systems, compare design variants, and investigate alignment mechanisms in future work.

\bibliographystyle{ACM-Reference-Format}
\bibliography{references2, references, related_work, references3}

\appendix

\end{document}